\documentclass[conference]{IEEEtran}
\IEEEoverridecommandlockouts
\usepackage{balance}
\usepackage{cite}
\usepackage{amsmath,amssymb,amsfonts}
\usepackage[ruled,linesnumbered]{algorithm2e}
\usepackage{graphicx}
\usepackage{booktabs}
\usepackage{diagbox}
\usepackage{pifont}
\usepackage{multirow}
\usepackage{textcomp}
\usepackage{xcolor}

\def\BibTeX{{\rm B\kern-.05em{\sc i\kern-.025em b}\kern-.08em
    T\kern-.1667em\lower.7ex\hbox{E}\kern-.125emX}}

\makeatletter
\algocf@newcmdside@kobe{LetIn@EKD}{%
    \KwSty{Ensemble Knowledge Distillation:}%
    \ifArgumentEmpty{#1}\relax{ #1}%
    \algocf@block{#2}{end}{#3}%
    \par
}

\newcommand\LetIn[2]{%
    \LetIn@EKD{#1}%
}
\makeatother
    
\begin{document}

\title{
Effective Intrusion Detection in Heterogeneous Internet-of-Things Networks via Ensemble Knowledge Distillation-based Federated Learning
}
\author{\IEEEauthorblockN{Jiyuan Shen\textsuperscript{1}, 
Wenzhuo Yang\textsuperscript{1}\textsuperscript{\textasteriskcentered}, Zhaowei Chu\textsuperscript{2}, Jiani Fan\textsuperscript{2}, Dusit Niyato\textsuperscript{2}, Kwok-Yan Lam\textsuperscript{1,2}\textsuperscript{\textasteriskcentered}}
\IEEEauthorblockA{\textsuperscript{1}Strategic Centre for Research in Privacy-Preserving Technologies \& Systems, Nanyang Technological University, Singapore\\
\textsuperscript{2}School of Computer Science and Engineering, Nanyang Technological University, Singapore\\
\{jiyuan001, zhaowei001, jiani001\}@e.ntu.edu.sg, \{wenzhuo.yang, dniyato, kwokyan.lam\}@ntu.edu.sg}}


\maketitle

\begin{abstract}

With the rapid development of low-cost consumer electronics and cloud computing, Internet-of-Things (IoT) devices are widely adopted for supporting next-generation distributed systems such as smart cities and industrial control systems. IoT devices are often susceptible to cyber attacks due to their open deployment environment and limited computing capabilities for stringent security controls. Hence, Intrusion Detection Systems (IDS) have emerged as one of the effective ways of securing IoT networks by monitoring and detecting abnormal activities. However, existing IDS approaches rely on centralized servers to generate behaviour profiles and detect anomalies, causing high response time and large operational costs due to communication overhead. Besides, sharing of behaviour data in an open and distributed IoT network environment may violate on-device privacy requirements. Additionally, various IoT devices tend to capture heterogeneous data, which complicates the training of behaviour models. In this paper, we introduce Federated Learning (FL) to collaboratively train a decentralized shared model of IDS, without exposing training data to others. Furthermore, we propose an effective method called Federated Learning Ensemble Knowledge Distillation (FLEKD) to mitigate the heterogeneity problems across various clients. FLEKD enables a more flexible aggregation method than conventional model fusion techniques. Experiment results on the public dataset CICIDS2019 demonstrate that the proposed approach outperforms local training and traditional FL in terms of both speed and performance and significantly improves the system's ability to detect unknown attacks. Finally, we evaluate our proposed framework's performance in three potential real-world scenarios and show FLEKD has a clear advantage in experimental results.

\end{abstract}


\section{Introduction}

\begin{figure}[ht]
    \centering
    \includegraphics[width=\linewidth]{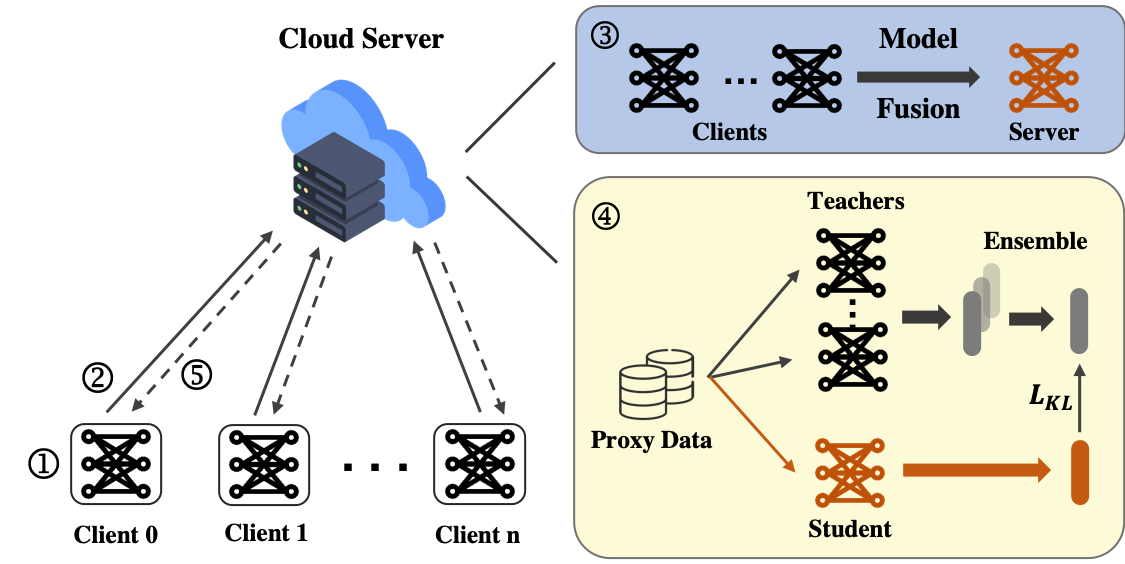}
    \caption{The overview of the FLEKD-IDS framework. \ding{172}: Train local model. \ding{173}: Transmit local models to server. \ding{174}: Fuse clients' models. \ding{175}: Ensemble knowledge distillation and fine-tune the global model. \ding{176}: Distribute the latest global model.}
    \label{fig:framework}
\end{figure}

In the era of digitalization, our daily life is gradually changed by ubiquitous Internet of Things (IoT) devices. These devices record and upload our personal information from day to night, thus generating exponential data every day for every person. 
By integrating with big data and edge computing devices, a lot of promising applications have emerged, such as intelligent industrial control systems and smart cities. 
Due to the openness of the IoT network deployment environment and the weak security control protocol operation ability caused by limited computing capabilities, IoT devices are very vulnerable to network attacks.
Attackers may comprise edge devices or databases through the network to steal or tamper with data, causing great social and economic impact~\cite{mukherjee1994network,liao2013intrusion}. 
Hence, security monitor detection measures should be considered to enhance IoT security.

As one of the most important security posture monitoring tools, the Intrusion Detection System (IDS) has attracted much attention in the context of IoT \cite{zarpelao2017survey}.
IDS can notify an administrator of potential cyber attacks by analyzing network traffic packets or system logs.  
Traditional IDS systems rely on rule-based approaches~\cite{han2003detecting} that are not effective to detect novel attacks. Furthermore, big data is another challenge for the rule-based IDSs. With the growth of IoT networks and increasing complexity of attacks, Machine Learning (ML) and Deep Learning (DL) methods have emerged as effective techniques for constructing automatic IDSs ~\cite{mishra2018detailed,ahmad2021network}. These techniques enable the system to learn from vast amounts of behaviour data, establish specific models efficiently, and identify attacking patterns that are difficult to detect manually.

However, the use of ML and DL in IDS for IoT networks comes with its own set of challenges. 
First, training models by local devices themselves will lead to poor performance because of the constrained computation ability and insufficient data.
Second, the communication overhead is large and will cause privacy concerns to transferring raw data of all devices to a central server for collaborative training.
Another challenge is handling the heterogeneous data collected by different devices~\cite{agrawal2022federated}. The data distribution, number of samples, and collection time on each edge device may be different. Besides, due to system updates or different device functions, the feature dimensions and attack types of data collected from different devices/clients may also be inconsistent. This heterogeneity further complicates the centralized ML and DL models for IoT security detection.

To address the first two concerns, we utilize an on-device privacy-preserving distributed machine learning paradigm, Federated Learning (FL). 
Instead of sharing local privacy datasets directly, FL can collaboratively train a global IDS model by transmitting local model parameters. Then, the central server aggregates local model parameters into a global one. Finally, the central server will dispatch the updated parameters to all local clients. The model will stop after the specified number of such communication synchronization rounds or after the model converges.

To mitigate the negative impacts, including slow model convergence speed, poor model prediction ability, etc, brought by data heterogeneity on diverse devices, we propose an effective heterogeneous federated learning framework using ensemble knowledge distillation (FLEKD), as shown in Figure~\ref{fig:framework}. Our scheme leverages mutually exclusive and unlabeled data to aggregate and transfer knowledge from all received heterogeneous client models toward a global model. This ensemble knowledge distillation technique allows a flexible aggregation method that can reduce the impact of clients' heterogeneity while not disclosing users' privacy and also can accelerate the model convergence speed.
Empirical experiment results demonstrated that our proposed framework can not only achieve accurate and timely intrusion detection but also narrow the knowledge gaps among heterogeneous clients. Our main contributions are summarized as follows:
\begin{itemize}
    \item We introduce an FL framework to develop an on-device collaborative deep intrusion detection model for edge devices in IoT networks.
    \item We propose a dynamic weight ensemble knowledge distillation scheme (FLEKD) to assist in mitigating the negative influence of clients' heterogeneity without violating users' personal privacy.
    \item We conduct extensive experiments on the public dataset CICIDS2019 to demonstrate better detection performance and improved ability to identify unknown attacks over local training models and naive FL global models.
    \item We assess the performance of our proposed framework in three possible real-world scenarios, namely, diverse data features, sample quantity, and missing certain classes. Our experimental results demonstrate that FLEKD exhibits clear and strong advantages.
\end{itemize}

\section{Background and Related Work}

\subsection{Intrusion Detection System}
The Intrusion Detection System (IDS) is a critical cybersecurity tool that plays a vital role in safeguarding the security and integrity of networked systems. By monitoring network traffic packets or system logs, IDS can identify anomalous behaviour or patterns that may indicate unauthorized use, misuse, or abuse \cite{mukherjee1994network} and alert system administrators of potential security breaches in real-time.
IDS also performs critical impact in the context of IoT security~\cite{fan2023understanding}. 
Traditional IDSs are rule-based or signature-based, which are effective for detecting known threats~\cite{han2003detecting}. 
However, they may not be as advanced as ML/DL-based IDSs in analyzing large volumes of data in an efficient manner and detecting emerging attacks \cite{mishra2018detailed,ahmad2021network}.
In recent years, there are also research efforts investigating emerging ML or DL paradigms to improve specific performance of IDS models, such as improving model adaptability to handle heterogeneous data by collaborative learning \cite{ma2023adcl} and reducing data annotation pressure by weakly supervised learning \cite{yang2021effective}.


\subsection{Federated Learning for Intrusion Detection}
Federated learning is designed for collaborative training of aggregated models with multiple devices or nodes without sharing raw data \cite{mcmahan2017communication}.
FL is a distributed paradigm that has the following advantages over centralized ML and DL models:
\begin{itemize}
    \item Keep raw data in the owner: In FL, devices/clients can maintain data privacy by training local models without sharing their own data with the central server. 
    \item Mitigate data scarcity: FL can access a much larger and more diverse dataset, which can help overcome issues related to limited data on some devices. The generalization ability of the aggregated model and the prediction ability of unknown items for the clients will be improved.
    \item Train with heterogeneous data: FL makes it possible to train models with large quantities of diverse data (features, sample quantity, and distribution, as well as the types of data that may be different) from multiple sources.   
\end{itemize}

FL has demonstrated its usefulness in the development of models to combat unknown attacks \cite{popoola2021federated}, achieve efficient anomaly detection \cite{liu2020communication,mothukuri2021federated}, and realize privacy-enhancing FL-based intrusion detection \cite{li2020deepfed, ruzafa2021intrusion}.
Moreover, FL has contributed to the development of various other IDSs, highlighting its potential as a valuable approach in the field of cybersecurity and IoT security \cite{agrawal2022federated,campos2022evaluating}.

A practical challenge of FL models is to maintain high performance when dealing with heterogeneous data collected from different devices. However, a limited number of FL-IDS models have comprehensively considered this issue.
Local model heterogeneity brought about by heterogeneous data will cause these problems: 1) slow convergence, 2) bias in the model, and 3) poor model accuracy. Therefore, in this paper, we propose an effective and efficient FL framework to address the data heterogeneity (diverse feature dimensions, sample quantity and distribution, and different types of data) for IDS in IoT security.

\section{An Effective Heterogeneous Intrusion Detection System Framework}

\subsection{Knowledge Distillation}
Knowledge distillation, first proposed by Hinton~\cite{KD}, allows transferring the knowledge of a large, complex model (known as the teacher) to a smaller, simpler model (known as the student). The motivation behind knowledge distillation is to train the student model to mimic the behaviour of the teacher model. The process of knowledge transfer usually needs a proxy dataset as the medium. As a result, most of the works will choose a mutually exclusive dataset while some others may use an autoencoder or GAN to generate some synthetic dataset. Once the proxy dataset is chosen, the class probabilities of the model's last layer (logits) or feature representations of middle hidden layers (feature maps) are usually used as the soft targets for knowledge transfer. The reason is that they contain more valuable information than the hard labels used in normal training.

For the original knowledge distillation, loss typically consists of two terms: a standard cross-entropy loss term and a distillation loss term. The former uses the hard label as the target while the latter uses a soft target. For the federated learning scenario, we usually do not have the publicly labelled dataset. Therefore, the combined loss is reduced to only the distillation loss term. The distillation loss term is often formulated as the Kullback-Leibler (KL) divergence between the teacher and student's softmax output probabilities, shown in Equation \eqref{eq:navie_kd}. 

\begin{equation}
\begin{aligned}
\label{eq:navie_kd}
    \sigma(z_{i}) &=\frac{\exp \left(z_{i} / T\right)}{\sum_{j} \exp \left(z_{j} / T\right)} \\
    L_{KL}(S \| T) &= \sum S_{i}(\sigma(z)) \log \left(\frac{S_{i}(\sigma(z))}{T_{i}(\sigma(z))}\right)
\end{aligned}
\end{equation}
where $S$ and $T$ represent the student and teacher logits respectively. $\sigma$ is the softmax function and a temperate $T$
is added. 

\begin{algorithm*}
\small
\caption{Federated Learning with Ensemble Knowledge Distillation}
\label{alg:FL_KD}
\KwIn{Proxy IDS dataset (unlabeled) $\mathcal{D}_0$, private IDS datasets (labeled) $D_K$, initialize clients models $\mathcal{W}_i$, the number of data points per client $n_i$.} 
\KwOut{Trained server model $\mathcal{W}_G$}
\For{each communication round $t=1$ to $T$}{
    Select a subset of clients $C_t$ to participate in the round\;
    \For{each client $i$ in $C_t$}{
        \textbf{Synchronize} the current global model $\mathcal{W}_G$ to client $i$\;        
        \textbf{Update} a local model $\mathcal{W}_i$ using client $i$'s private dataset $D_K$\;        
        \textbf{Transmit} the client models $\mathcal{W}_i$ to the central server\;
    }
    \textbf{Model Fusion:} The server computes an updated consensus, which is an average of client models parameters 
    $\mathcal{W}_G = \sum_{i \in C_t} \frac{n_{i}}{\sum_{i \in C_{t}} n_{i}} W_i$\;
    \LetIn
        {Student: Calculate the logit vectors of server model $\mathbf{x}_t^s$ based on the proxy dataset $\mathcal{D}_0$\;
    Teacher: Calculate the logit vectors of client models based on the proxy dataset $\mathcal{D}_0$. Use ensemble algorithm based on Equation~\eqref{weight_client} to get the final teacher knowledge $\mathbf{\hat{x}}_t^k$\;
    \textbf{Fine-tune} the global model $\mathcal{W}_G$ using Kullback-Leibler divergence $\mathcal{L} = KL(\mathbf{x}_t^s, \mathbf{\hat{x}}_t^k)$\;}

}
Return $\mathcal{W}_G$
\end{algorithm*}

\subsection{FLEKD: Using Ensemble Knowledge Distillation for IDS in Heterogeneous IoT Networks}

We introduce federated learning as a privacy-preserving collaborative training paradigm for the IDS model. The conventional FL algorithm commonly involves three steps: First, synchronizing the current global model parameters to maintain consistency among each client. Second, updating the local model parameters on private data using Adam or SGD as optimizer. Third, transmitting the models of each client to the server side and integrating them by using a specific aggregation algorithm in Equation \eqref{eq:agg}. 
\begin{equation}
\begin{aligned}
\label{eq:agg}
    W_{t+1} = W_{t}+ \frac{1}{C_t} \sum_{i=1}^{C_t} F_{t+1}^{i}
\end{aligned}
\end{equation}
where $\frac{1}{C_t} \sum_{i=1}^{C_t} F_{t+1}^{i}$ denotes the average aggregation of client models $F_{t+1}^{i}$. These three steps constitute a loop until the global model converges, as shown in steps 3-8 of Algorithm~\ref{alg:FL_KD}.

Usually, average or weighted average algorithms are used for server-side aggregation. However, many research works~\cite{zhao2018federated,li2019convergence} suggest that this may not be the best aggregation mode, especially on heterogeneous data. In fact, simple averaging algorithms result in the loss of a lot of useful information from client models. Specifically, due to the imbalance in data distribution, some clients may be better at detecting certain attack patterns but not others. After aggregation, it is highly likely that the classification boundaries that were originally clear for certain categories become fuzzy, which affects the overall detection performance~\cite{ensemble}. Therefore, this work aims to better utilize the effective information from each client to further improve the generalization and performance of the server model after aggregation.

Our proposed FLEKD method, shown in Figure~\ref{fig:framework}, can effectively utilize heterogeneous IDS data. To be specific, we combine the idea of one-to-many knowledge distillation with federated learning, using the server-side model obtained by simple aggregation as the student model, and client models as the teacher models. The student model acquires knowledge from logits ensembled overall the received teacher models, and thus mutually beneficial information can be shared. In the next round, each activated client receives the corresponding fused prototype model.

In one-to-many knowledge distillation, we propose to use an ensemble method to compress the knowledge of the teacher model. Specifically, we propose a dynamic weight ensemble method. We first test each client model to obtain its score on the test set. Then, we perform a Softmax operation on the test scores, with the addition of a deterministic temperature to enlarge the difference between clients and increase the teacher knowledge's reliance on higher-scoring clients. Therefore, the dynamic weight for each client at this round is generated. Finally, we perform a matrix multiplication between the dynamic weights and the logits from each client, shown in Equation \eqref{weight_client}, to obtain the final ensemble knowledge. 

\begin{equation}
\begin{aligned}
    \label{weight_client}
    Client_{i}&=\frac{\exp \left(acc_{i} / DT\right)}{\sum_{j} \exp \left(acc_{j} / DT\right)} \\
    EKD &= Client \times logits
\end{aligned}
\end{equation}

After acquiring the ensemble knowledge distillation, similar to the original KD, we use KL divergence to fine-tune the server-side model, as detailed in steps 9-13 of Algorithm~\ref{alg:FL_KD}.

Notably, as federated learning requires each client to upload its own model to the server, our proposed method does not impose additional burdens or communication overhead.

\section{Experiments}
\subsection{Experiment Setup}
\textbf{Dataset and model.} We perform our experiments on the CICIDS2019 dataset. It contains normal and the latest common DDoS attack events, similar to real data (PCAPs). It also includes network traffic analysis results using CICFlowMeter-V3, which is based on timestamp, source and destination IP, source and destination port, protocol and attacking tag stream. According to the feature dimensions of CICIDS2018 and 2017, we split the dataset into three groups, clients 0-2 with 82 full feature dimensions, clients 3-5 with 79 dimensions, and clients 6-8 with only 24 dimensions. Note that the number of UDPLag attack samples in the original CICIDS2019 dataset is extremely small, about twenty-third of the number of other categories. 

For the model selection, we use the naive MLP neural network directly. Our network contains three layers and the hidden neuron is set to 128. We use ReLU as the activation function. All the clients' model is initialized with the same random parameters.

\addtolength{\topmargin}{0.01in}

\textbf{Heterogeneous distribution of client data.} To simulate a realistic and practical environment, we introduce heterogeneity in the distribution of client data. We adopt the use of the Dirichlet distribution, as proposed in previous studies such as~\cite{yurochkin2019bayesian}. In this approach, we create disjoint non-i.i.d. client training data. The parameter $\alpha$ is used to control the degree of non-i.i.d. For example, setting $\alpha$ to 100 can mimic identical local data distributions, while smaller $\alpha$ values result in more heterogeneous data distributions.

\textbf{Parameters setting.} In the federated learning process, the client number is set to 9. According to the client number, every three clients have private data sets of the same dimension. We conduct two rounds of local training before the communication with the server side. We use Adam as the optimization and set the learning rate to 0.001 with a uniform decay strategy of five per cent. In the ensemble knowledge distillation process, we first need to generate a proxy dataset. The proxy dataset contains 1260 samples. We adopted a uniform sampling method, so the samples contained in each category except UDPLag are basically consistent, while the UDPLag is about a third of the number of other attacks. We set the temperature at 1.5 for knowledge distillation. For the ensembling weights, we set the deterministic temperature to 0.5 to enlarge the discrepancy of each client. Similarly, we also use Adam as the optimization and set the learning rate to 0.001 with a decay strategy of five per cent for each epoch.

\begin{table*}[ht]
\renewcommand\arraystretch{1.6}
\Huge
\begin{center}
\caption{Main results. Clients 0-8 conduct local training and all the scores are calculated by F1-score.}
\label{tab:main}
\resizebox{\linewidth}{!}{
\begin{tabular}{@{}cccc|ccc|ccc|ccc|ccc|ccc|ccc@{}}
\toprule
    \multicolumn{1}{c}{\multirow{2}{*}{\begin{tabular}[c]{@{}c@{}}\textbf{Local}\\ \textbf{Training}\end{tabular}}} 
                  & \multicolumn{3}{c}{\textbf{Portmap}}                                     & \multicolumn{3}{c}{\textbf{LDAP}}                                        & \multicolumn{3}{c}{\textbf{MSSQL}}                                       & \multicolumn{3}{c}{\textbf{NetBIOS}}                                     & \multicolumn{3}{c}{\textbf{Syn}}                                         & \multicolumn{3}{c}{\textbf{UDP}}                                         & \multicolumn{3}{c}{\textbf{UDPLag}}                                      \\ \cmidrule{2-22}
    & \multicolumn{1}{c}{$\alpha=$10} & \multicolumn{1}{c}{1} & \multicolumn{1}{c}{0.5} & \multicolumn{1}{c}{10} & \multicolumn{1}{c}{1} & \multicolumn{1}{c}{0.5} & \multicolumn{1}{c}{10} & \multicolumn{1}{c}{1} & \multicolumn{1}{c}{0.5} & \multicolumn{1}{c}{10} & \multicolumn{1}{c}{1} & \multicolumn{1}{c}{0.5} & \multicolumn{1}{c}{10} & \multicolumn{1}{c}{1} & \multicolumn{1}{c}{0.5} & \multicolumn{1}{c}{10} & \multicolumn{1}{c}{1} & \multicolumn{1}{c}{0.5} & \multicolumn{1}{c}{10} & \multicolumn{1}{c}{1} & \multicolumn{1}{c}{0.5} \\ \midrule

\multicolumn{1}{c|}{\textbf{Client 0}} & 99.89\%                & 99.74\%               & 70.08\%                 & 99.83\%                & 99.78\%               & 0.00\%                  & 99.96\%                & 99.85\%               & 99.97\%                 & 99.84\%                & 99.69\%               & 99.54\%                 & 99.91\%                & 99.90\%               & 99.92\%                 & 99.98\%                & 99.96\%               & 92.73\%                 & 97.27\%                & 96.77\%               & 97.99\%                 \\
\multicolumn{1}{c|}{\textbf{Client 1}} & 99.88\%                & 99.89\%               & 99.51\%                 & 99.86\%                & 99.80\%               & 96.59\%                 & 99.96\%                & 99.97\%               & 95.78\%                 & 99.88\%                & 99.83\%               & 99.72\%                 & 99.93\%                & 99.94\%               & 98.78\%                 & 99.98\%                & 99.92\%               & 99.96\%                 & 97.52\%                & 98.42\%               & 0.00\%                  \\
\multicolumn{1}{c|}{\textbf{Client 2}} & 99.92\%                & 99.82\%               & 99.66\%                 & 99.94\%                & 99.92\%               & 99.83\%                 & 99.93\%                & 99.93\%               & 99.71\%                 & 99.94\%                & 99.87\%               & 99.81\%                 & 99.93\%                & 99.90\%               & 99.87\%                 & 99.99\%                & 100.00\%              & 99.46\%                 & 97.30\%                & 97.03\%               & 76.48\%                 \\
\multicolumn{1}{c|}{\textbf{Client 3}} & 99.81\%                & 99.60\%               & 98.96\%                 & 99.17\%                & 99.76\%               & 96.76\%                 & 99.89\%                & 99.95\%               & 99.92\%                 & 99.80\%                & 99.51\%               & 98.70\%                 & 99.92\%                & 99.89\%               & 99.93\%                 & 99.29\%                & 99.90\%               & 99.94\%                 & 97.01\%                & 97.11\%               & 56.63\%                 \\
\multicolumn{1}{c|}{\textbf{Client 4}} & 99.84\%                & 2.30\%                & 64.93\%                 & 99.92\%                & 99.86\%               & 99.77\%                 & 99.90\%                & 66.47\%               & 94.97\%                 & 99.89\%                & 77.55\%               & 0.00\%                  & 99.92\%                & 99.90\%               & 99.11\%                 & 99.99\%                & 99.86\%               & 99.73\%                 & 97.40\%                & 94.12\%               & 67.13\%                 \\
\multicolumn{1}{c|}{\textbf{Client 5}} & 99.83\%                & 99.97\%               & 98.91\%                 & 99.90\%                & 99.90\%               & 83.64\%                 & 99.93\%                & 99.98\%               & 98.77\%                 & 99.89\%                & 99.90\%               & 99.52\%                 & 99.87\%                & 99.92\%               & 99.86\%                 & 99.96\%                & 99.99\%               & 75.67\%                 & 97.03\%                & 97.63\%               & 89.07\%                 \\
\multicolumn{1}{c|}{\textbf{Client 6}} & 90.65\%                & 95.13\%               & 8.94\%                  & 99.78\%                & 99.84\%               & 99.29\%                 & 80.23\%                & 67.99\%               & 2.42\%                  & 93.57\%                & 39.35\%               & 52.28\%                 & 99.83\%                & 99.84\%               & 98.95\%                 & 99.99\%                & 99.99\%               & 99.91\%                 & 93.55\%                & 93.28\%               & 46.65\%                 \\
\multicolumn{1}{c|}{\textbf{Client 7}} & 94.86\%                & 14.50\%               & 32.67\%                 & 99.84\%                & 99.89\%               & 99.71\%                 & 89.36\%                & 21.53\%               & 13.57\%                 & 95.19\%                & 70.78\%               & 53.99\%                 & 99.81\%                & 99.90\%               & 99.05\%                 & 99.91\%                & 99.99\%               & 99.99\%                 & 94.22\%                & 93.05\%               & 54.84\%                 \\
\multicolumn{1}{c|}{\textbf{Client 8}} & 97.19\%                & 91.88\%               & 96.53\%                 & 99.82\%                & 99.58\%               & 94.26\%                 & 93.11\%                & 84.77\%               & 0.00\%                  & 95.94\%                & 95.06\%               & 68.50\%                 & 99.89\%                & 99.92\%               & 99.82\%                 & 99.94\%                & 99.90\%               & 97.11\%                 & 93.31\%                & 95.59\%               & 49.36\%                 \\ \midrule
\multicolumn{1}{c}{\multirow{2}{*}{\begin{tabular}[c]{@{}c@{}}\textbf{Distributed}\\ \textbf{Training}\end{tabular}}} 
 & \multicolumn{3}{c}{\textbf{Portmap}}                                     & \multicolumn{3}{c}{\textbf{LDAP}}                                        & \multicolumn{3}{c}{\textbf{MSSQL}}                                       & \multicolumn{3}{c}{\textbf{NetBIOS}}                                     & \multicolumn{3}{c}{\textbf{Syn}}                                         & \multicolumn{3}{c}{\textbf{UDP}}                                         & \multicolumn{3}{c}{\textbf{UDPLag}}                                      \\ \cmidrule{2-22}
    & \multicolumn{1}{c}{$\alpha=$10} & \multicolumn{1}{c}{1} & \multicolumn{1}{c}{0.5} & \multicolumn{1}{c}{10} & \multicolumn{1}{c}{1} & \multicolumn{1}{c}{0.5} & \multicolumn{1}{c}{10} & \multicolumn{1}{c}{1} & \multicolumn{1}{c}{0.5} & \multicolumn{1}{c}{10} & \multicolumn{1}{c}{1} & \multicolumn{1}{c}{0.5} & \multicolumn{1}{c}{10} & \multicolumn{1}{c}{1} & \multicolumn{1}{c}{0.5} & \multicolumn{1}{c}{10} & \multicolumn{1}{c}{1} & \multicolumn{1}{c}{0.5} & \multicolumn{1}{c}{10} & \multicolumn{1}{c}{1} & \multicolumn{1}{c}{0.5} \\ \midrule
\multicolumn{1}{c|}{\textbf{FL}}       & 99.00\%                & 99.45\%               & 99.54\%                 & 99.67\%                & 99.64\%               & 99.83\%                 & 99.98\%                & 99.86\%               & 99.62\%                 & 98.97\%                & 99.57\%               & 99.73\%                 & 99.93\%                & 99.93\%               & 99.88\%                 & 99.87\%                & 99.78\%               & 99.99\%                 & 97.55\%                & 97.21\%               & 95.40\%                 \\
\multicolumn{1}{c|}{\textbf{FLEKD}}     & \textbf{99.80\%}       & \textbf{99.73\%}      & \textbf{99.77\%}        & \textbf{99.84\%}       & \textbf{99.70\%}      & \textbf{99.91\%}        & \textbf{99.99\%}       & \textbf{99.98\%}      & \textbf{99.90\%}        & \textbf{99.82\%}       & \textbf{99.73\%}      & \textbf{99.86\%}        & \textbf{99.94\%}       & \textbf{99.93\%}      & \textbf{99.92\%}        & \textbf{99.91\%}       & \textbf{99.88\%}      & \textbf{99.99\%}        & \textbf{98.34\%}       & \textbf{97.64\%}      & \textbf{96.62\%}        \\ \bottomrule
\end{tabular}
}
\end{center}
\end{table*}

\subsection{Performance of Proposed Framework}
In Table~\ref{tab:main}, we present our main result. The table shows the performance of our model using three different sample parameters for the Dirichlet distribution: $\alpha$ = 10, 1, and 0.5. The first part of the table displays the performance of the local training process conducted by each client on their private dataset.
We observe that as the data becomes more heterogeneous, there is a significant decrease in the F1-score of each client. In fact, when $\alpha$ is set to 0.5, some clients exhibit completely incorrect predictions for certain classes. This is due to the non-i.i.d. nature of the private datasets, which may result in some clients having an imbalanced amount of data and experiencing underfitting. This can lead to a decrease in the prediction accuracy for other classes as well. Therefore, when the problem of imbalanced data occurs in on-device local training, it can easily lead to a significant decrease in the IoT device's prediction performance.

\begin{figure}[h]
    \centering
    \includegraphics[width=0.8\linewidth]{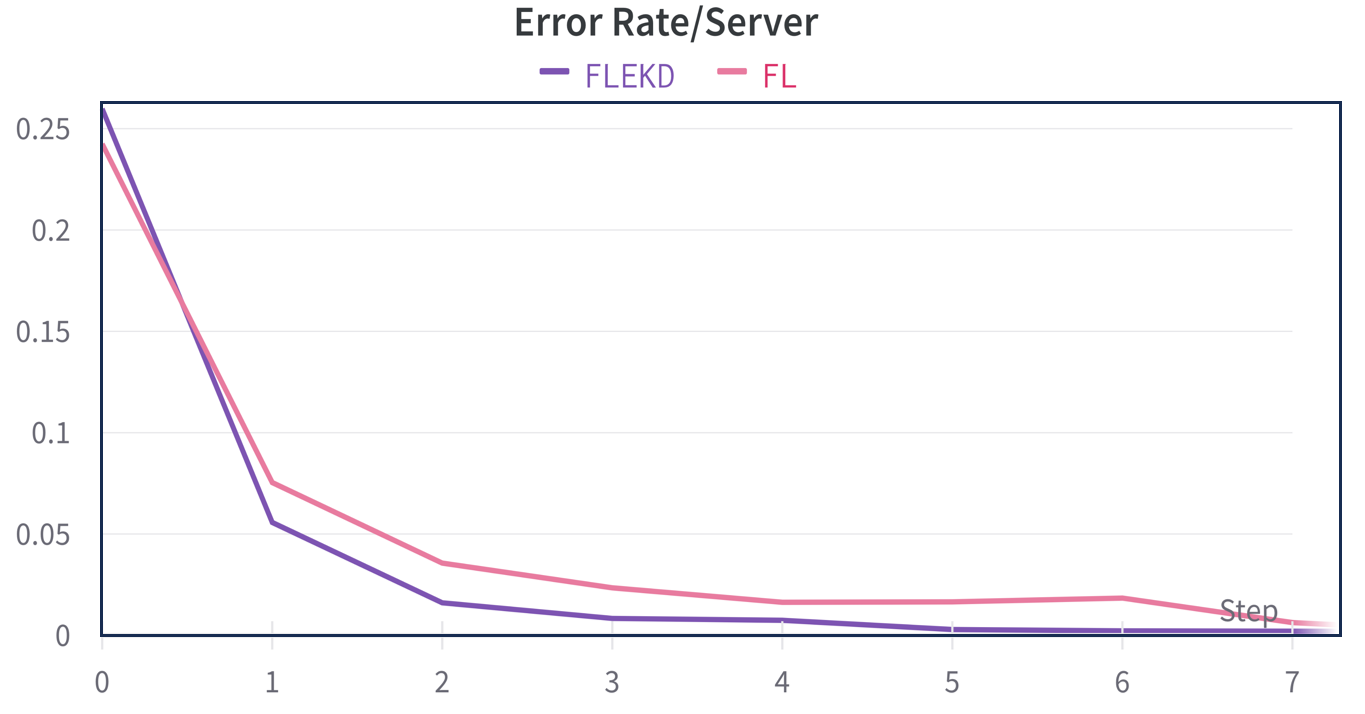}
    \setlength{\abovecaptionskip}{-0.3cm}
    \caption{Error rate of the server-side model on the test set.}
    \label{fig:converge}
\end{figure}

However, by introducing distributed training, as shown in the second part of Table~\ref{tab:main}, the predictive performance can be improved significantly. For example, in the case of UDPLag, a class with very few examples, an F1-score of 95.40\% can be achieved with $\alpha$ set to 0.5. This is a large improvement compared to the results obtained from local training, where clients 1, 6, and 8 achieved less than 50\% accuracy. Furthermore, F1-score improves in various other intrusion attack scenarios. Our proposed EKD fine-tuning further enhances the performance of our central model and accelerates the convergence rate. For each setting, the number of communication rounds required to achieve a 95\% F1-score was reduced by one when using the FLEKD method. This improved efficiency is further illustrated in Figure~\ref{fig:converge}, where the convergence speed of FLEKD is noticeably faster than that of the original FL under the same conditions, allowing for faster achievement of optimal final results, which demonstrates the effectiveness of using FLEKD.

\subsection{Analysis}
In this paper, we focus on evaluating the effectiveness of our proposed method in the context of an intrusion detection system in IoT security. To this end, we consider three key scenarios that are likely to arise in practice and investigate the performance of our method under each of them. Specifically, we examine the impact of (1) varying dimensions of the data, (2) different sample sizes, and (3) different attack category distributions in various clients. Through our experiments, we demonstrate the superiority of our approach in these scenarios.

\textbf{The impact of different dimensions.}
We consider the possibility that data collected from different time periods may have varying dimensions, for instance, the CICIDS2017 dataset contains only 24 feature dimensions compared to CICIDS2019 with 82 dimensions. In real-world application scenarios, it is necessary to combine data collected from different IoT devices for training. We present three scenarios in Table~\ref{tab:dim}, where we compare the results of training with datasets of different dimensions locally and with federated learning. Our results show that when the available feature vectors are reduced to 24 dimensions, there is a significant decrease in performance, with F1-score dropping from the original 96.02\% to 84.71\% using only local training. However, when we apply federated learning and our proposed method of ensemble knowledge distillation, we achieve a significant improvement in accuracy, with a score of 99.80\%. This indicates that FLEKD can effectively solve the problem of dimensionality differences in IoT device data.

\begin{table}[t]
\small
\centering
\caption{The impact of different dimensions.}
\label{tab:dim}
\resizebox{0.6\linewidth}{!}{%
\begin{tabular}{@{}cccc@{}}
\toprule
       & Precision        & Recall           & F1-score         \\ \midrule
82 dim & 95.93\%          & 96.39\%          & 96.02\%          \\
79 dim & 86.31\%          & 86.75\%          & 86.17\%          \\
24 dim & 84.17\%          & 85.48\%          & 84.71\%          \\ \midrule
FL     & 99.37\%          & 99.36\%          & 99.37\%          \\
FLEKD   & \textbf{99.80\%} & \textbf{99.80\%} & \textbf{99.80\%} \\ \bottomrule
\end{tabular}%
}
\end{table}
\vspace{0.5cm}

\textbf{The impact of different sample sizes.}
Due to the characteristics of network attacks, which may occur in short periods with high frequency and target certain vulnerable devices, there may be significant differences in the number of samples between IoT devices. Therefore, we divided the dataset into three groups of clients with different sample sizes. Specifically, the number of samples in each group differs by a factor of ten, with the group containing the least number of samples represented as `Base'. In Table~\ref{tab:sample}, we observe a negative correlation between the number of samples and the detection performance when using local training only. We achieved the best performance of 99.50\% on 100$*$Base. However, due to the lack of communication between different clients, the base model was not able to benefit from the larger number of samples in 100$*$Base, and only achieved an F1-score of 84.71\%.
When we used FLEKD to aggregate and optimize the models of different clients, the final F1-score reached 99.80\%, which is higher than all results obtained by local training and ordinary FL. This demonstrates that FLEKD can effectively address the sample size differences between different IoT devices and significantly improve intrusion detection performance.

\begin{table}[t]
\small
\centering
\caption{The impact of different sample sizes.}
\label{tab:sample}
\resizebox{0.6\linewidth}{!}{%
\begin{tabular}{@{}cccc@{}}
\toprule
         & Precision        & Recall           & F1-score         \\ \midrule
Base     & 84.17\%          & 85.48\%          & 84.71\%          \\
10*Base  & 92.97\%          & 88.55\%          & 87.25\%          \\
100*Base & 99.51\%          & 99.50\%          & 99.50\%          \\ \midrule
FL       & 99.37\%          & 99.36\%          & 99.37\%          \\
FLEKD     & \textbf{99.80\%} & \textbf{99.80\%} & \textbf{99.80\%} \\ \bottomrule
\end{tabular}%
}
\end{table}

\textbf{The impact of different attack category distributions.}
In real-world scenarios, different IoT devices usually own heterogeneous data distributions.
It is common to encounter unknown attacks but traditional centralized IDSs unable to identify novel attacks effectively. Hence, in our experiments, we simulated the situation where each IoT device holds a different data distribution. There are seven classes of attacks in the CICIDS2019 dataset, so we set up seven local clients, each device missing a specific category of attack. As shown in Table~\ref{tab:drop}, it is evident that all clients failed to detect the lacking attack. However, the detection abilities for unknown attacks of the FL-based and FLEKD-based IDS models are high. The overall detection performance of the proposed method for all attacks is also significantly improved.
By adopting FLEKD, we find that even the most difficult-to-detect `UDPLag' can reach 80.86\%, while other corresponding attack categories can be improved to about 99\%. This demonstrates that FLEKD-based IDS can effectively deal with the category heterogeneity problem, enabling different devices to identify attacks they have not seen before.

\begin{table}[ht]
\renewcommand\arraystretch{1.4}
\centering
\caption{The impact of different attack category distributions.}
\label{tab:drop}
\resizebox{\linewidth}{!}{%
\begin{tabular}{@{}cccccccc@{}}
\toprule
             & 0: Portmap       & 1: LDAP          & 2:MSSQL          & 3:NetBIOS        & 4: Syn           & 5: UDP           & 6: UDPLag        \\ \midrule
Drop Label 0 & 0.00\%           & 98.72\%          & 59.09\%          & 95.62\%          & 99.67\%          & 95.10\%          & 0.00\%           \\
Drop Label 1 & 99.30\%          & 0.00\%           & 66.45\%          & 99.62\%          & 99.91\%          & 99.98\%          & 94.95\%          \\
Drop Label 2 & 66.77\%          & 99.96\%          & 0.00\%           & 99.75\%          & 99.95\%          & 99.99\%          & 96.81\%          \\
Drop Label 3 & 91.06\%          & 97.03\%          & 54.03\%          & 0.00\%           & 99.68\%          & 95.84\%          & 0.00\%           \\
Drop Label 4 & 75.76\%          & 99.69\%          & 94.54\%          & 95.72\%          & 0.00\%           & 99.90\%          & 11.05\%          \\
Drop Label 5 & 99.74\%          & 66.62\%          & 99.78\%          & 99.83\%          & 99.92\%          & 0.00\%           & 97.47\%          \\
Drop Label 6 & 91.21\%          & 97.93\%          & 0.00\%           & 72.35\%          & 99.65\%          & 94.46\%          & 0.00\%           \\ \midrule
FL           & 99.05\%          & 99.80\%          & 94.37\%          & 97.31\%          & 99.92\%          & 98.85\%          & 75.28\%          \\
FLEKD        & \textbf{99.66\%} & \textbf{99.88\%} & \textbf{98.97\%} & \textbf{99.80\%} & \textbf{99.94\%} & \textbf{99.97\%} & \textbf{80.86\%} \\ \bottomrule
\end{tabular}%
}
\end{table}

\vspace{-0.1cm}
\section{Conclusion}
In this work, we introduce federated learning towards IDS in IoT networks to address the privacy issue and central training problem. For non-i.i.d. situations, We propose FLEKD to mitigate the negative influence of clients’ heterogeneity without increasing additional communication overhead. Our proposed framework FLEKD outperforms the local training and original FL results in both speed and performance on the CICIDS2019 dataset. Besides, FLEKD exhibits clear superiority in handling heterogeneous data and achieves better results over three real-world scenarios. Overall, our work offers a promising solution for IoT networks' security.

\section{Acknowledge}
This research is supported by the National Research Foundation, Singapore under its Strategic Capability Research Centres Funding Initiative. Any opinions, findings and conclusions or recommendations expressed in this material are those of the author(s) and do not reflect the views of National Research Foundation, Singapore.

\balance
\tiny
\bibliographystyle{IEEEtran}
\bibliography{IEEEabrv,mybibfile}

\begin{thebibliography}{10}
\providecommand{\url}[1]{#1}
\csname url@samestyle\endcsname
\providecommand{\newblock}{\relax}
\providecommand{\bibinfo}[2]{#2}
\providecommand{\BIBentrySTDinterwordspacing}{\spaceskip=0pt\relax}
\providecommand{\BIBentryALTinterwordstretchfactor}{4}
\providecommand{\BIBentryALTinterwordspacing}{\spaceskip=\fontdimen2\font plus
\BIBentryALTinterwordstretchfactor\fontdimen3\font minus
  \fontdimen4\font\relax}
\providecommand{\BIBforeignlanguage}[2]{{%
\expandafter\ifx\csname l@#1\endcsname\relax
\typeout{** WARNING: IEEEtran.bst: No hyphenation pattern has been}%
\typeout{** loaded for the language `#1'. Using the pattern for}%
\typeout{** the default language instead.}%
\else
\language=\csname l@#1\endcsname
\fi
#2}}
\providecommand{\BIBdecl}{\relax}
\BIBdecl

\bibitem{mukherjee1994network}
B.~Mukherjee, L.~T. Heberlein, and K.~N. Levitt, ``Network intrusion
  detection,'' \emph{IEEE network}, vol.~8, no.~3, pp. 26--41, 1994.

\bibitem{liao2013intrusion}
H.-J. Liao, C.-H.~R. Lin, Y.-C. Lin, and K.-Y. Tung, ``Intrusion detection
  system: A comprehensive review,'' \emph{Journal of Network and Computer
  Applications}, vol.~36, no.~1, pp. 16--24, 2013.

\bibitem{zarpelao2017survey}
B.~B. Zarpel{\~a}o, R.~S. Miani, C.~T. Kawakani, and S.~C. de~Alvarenga, ``A
  survey of intrusion detection in internet of things,'' \emph{Journal of
  Network and Computer Applications}, vol.~84, pp. 25--37, 2017.

\bibitem{han2003detecting}
S.-J. Han and S.-B. Cho, ``Detecting intrusion with rule-based integration of
  multiple models,'' \emph{Computers \& Security}, vol.~22, no.~7, pp.
  613--623, 2003.

\bibitem{mishra2018detailed}
P.~Mishra, V.~Varadharajan, U.~Tupakula, and E.~S. Pilli, ``A detailed
  investigation and analysis of using machine learning techniques for intrusion
  detection,'' \emph{IEEE communications surveys \& tutorials}, vol.~21, no.~1,
  pp. 686--728, 2018.

\bibitem{ahmad2021network}
Z.~Ahmad, A.~Shahid~Khan, C.~Wai~Shiang, J.~Abdullah, and F.~Ahmad, ``Network
  intrusion detection system: A systematic study of machine learning and deep
  learning approaches,'' \emph{Transactions on Emerging Telecommunications
  Technologies}, vol.~32, no.~1, p. e4150, 2021.

\bibitem{agrawal2022federated}
S.~Agrawal, S.~Sarkar, O.~Aouedi, G.~Yenduri, K.~Piamrat, M.~Alazab,
  S.~Bhattacharya, P.~K.~R. Maddikunta, and T.~R. Gadekallu, ``Federated
  learning for intrusion detection system: Concepts, challenges and future
  directions,'' \emph{Computer Communications}, 2022.

\bibitem{fan2023understanding}
J.~Fan, W.~Yang, Z.~Liu, J.~Kang, D.~Niyato, K.-Y. Lam, and H.~Du,
  ``Understanding security in smart city domains from the ant-centric
  perspective,'' \emph{IEEE Internet of Things Journal}, 2023.

\bibitem{ma2023adcl}
Z.~Ma, L.~Liu, W.~Meng, X.~Luo, L.~Wang, and W.~Li, ``Adcl: Towards an adaptive
  network intrusion detection system using collaborative learning in iot
  networks,'' \emph{IEEE Internet of Things Journal}, 2023.

\bibitem{yang2021effective}
W.~Yang and K.-Y. Lam, ``Effective anomaly detection model training with only
  unlabeled data by weakly supervised learning techniques,'' in \emph{ICICS
  2021}, 2021, pp. 402--425.

\bibitem{mcmahan2017communication}
B.~McMahan, E.~Moore, D.~Ramage, S.~Hampson, and B.~A. y~Arcas,
  ``Communication-efficient learning of deep networks from decentralized
  data,'' in \emph{Artificial intelligence and statistics}.\hskip 1em plus
  0.5em minus 0.4em\relax PMLR, 2017, pp. 1273--1282.

\bibitem{popoola2021federated}
S.~I. Popoola, R.~Ande, B.~Adebisi, G.~Gui, M.~Hammoudeh, and O.~Jogunola,
  ``Federated deep learning for zero-day botnet attack detection in iot-edge
  devices,'' \emph{IEEE Internet of Things Journal}, vol.~9, no.~5, pp.
  3930--3944, 2021.

\bibitem{liu2020communication}
Y.~Liu, N.~Kumar, Z.~Xiong, W.~Y.~B. Lim, J.~Kang, and D.~Niyato,
  ``Communication-efficient federated learning for anomaly detection in
  industrial internet of things,'' in \emph{IEEE Global Communications
  Conference}.\hskip 1em plus 0.5em minus 0.4em\relax IEEE, 2020, pp. 1--6.

\bibitem{mothukuri2021federated}
V.~Mothukuri, P.~Khare, R.~M. Parizi, S.~Pouriyeh, A.~Dehghantanha, and
  G.~Srivastava, ``Federated-learning-based anomaly detection for iot security
  attacks,'' \emph{IEEE Internet of Things Journal}, vol.~9, no.~4, pp.
  2545--2554, 2021.

\bibitem{li2020deepfed}
B.~Li, Y.~Wu, J.~Song, R.~Lu, T.~Li, and L.~Zhao, ``Deepfed: Federated deep
  learning for intrusion detection in industrial cyber--physical systems,''
  \emph{IEEE Transactions on Industrial Informatics}, vol.~17, no.~8, pp.
  5615--5624, 2020.

\bibitem{ruzafa2021intrusion}
P.~Ruzafa-Alc{\'a}zar, P.~Fern{\'a}ndez-Saura, E.~M{\'a}rmol-Campos,
  A.~Gonz{\'a}lez-Vidal, J.~L. Hern{\'a}ndez-Ramos, J.~Bernal-Bernabe, and
  A.~F. Skarmeta, ``Intrusion detection based on privacy-preserving federated
  learning for the industrial iot,'' \emph{IEEE Transactions on Industrial
  Informatics}, vol.~19, no.~2, pp. 1145--1154, 2021.

\bibitem{campos2022evaluating}
E.~M. Campos, P.~F. Saura, A.~Gonz{\'a}lez-Vidal, J.~L. Hern{\'a}ndez-Ramos,
  J.~B. Bernab{\'e}, G.~Baldini, and A.~Skarmeta, ``Evaluating federated
  learning for intrusion detection in internet of things: Review and
  challenges,'' \emph{Computer Networks}, vol. 203, p. 108661, 2022.

\bibitem{KD}
G.~Hinton, O.~Vinyals, and J.~Dean, ``Distilling the knowledge in a neural
  network,'' \emph{arXiv preprint arXiv:1503.02531}, 2015.

\bibitem{zhao2018federated}
Y.~Zhao, M.~Li, L.~Lai, N.~Suda, D.~Civin, and V.~Chandra, ``Federated learning
  with non-iid data,'' \emph{arXiv preprint arXiv:1806.00582}, 2018.

\bibitem{li2019convergence}
X.~Li, K.~Huang, W.~Yang, S.~Wang, and Z.~Zhang, ``On the convergence of fedavg
  on non-iid data,'' \emph{arXiv preprint arXiv:1907.02189}, 2019.

\bibitem{ensemble}
T.~Lin, L.~Kong, S.~U. Stich, and M.~Jaggi, ``Ensemble distillation for robust
  model fusion in federated learning,'' \emph{Advances in Neural Information
  Processing Systems}, vol.~33, pp. 2351--2363, 2020.

\bibitem{yurochkin2019bayesian}
M.~Yurochkin, M.~Agarwal, S.~Ghosh, K.~Greenewald, N.~Hoang, and Y.~Khazaeni,
  ``Bayesian nonparametric federated learning of neural networks,'' in
  \emph{International conference on machine learning}.\hskip 1em plus 0.5em
  minus 0.4em\relax PMLR, 2019, pp. 7252--7261.

\end{thebibliography}
\end{document}